KonanAI LLC

# aoip.ai: An Open-Source P2P SDK Enabling Decentralized, Transparent, & Democratized AI in VoIP and IoT

A White Paper

December 2023

Joseph Konan, Shikhar Agnihotri, and Chia-Chun Hsieh

**Executive Summary:** This white paper introduces aoip.ai, a groundbreaking open-source SDK incorporating peer-to-peer technology and advanced AI integration to transform VoIP and IoT applications. It addresses key market challenges by enhancing data security, elevating communication quality, and providing greater flexibility for developers and users. Developed in collaboration with Carnegie Mellon University, aoip.ai sets a new standard for decentralized and democratized communication solutions.

## Preface

In an era marked by the rapid evolution of communication technologies, a pioneering breakthrough emerges: aoip.ai. This novel innovation reshapes the Voice over Internet Protocol (VoIP) and Internet of Things (IoT) landscape, offering an advanced open-source Software Development Kit (SDK) infused with cutting-edge audio AI capabilities. At its core, aoip.ai addresses prevalent challenges in the current market, setting new benchmarks in connectivity, transparency, and technological empowerment.

*Peer-To-Peer Decentralization:* Central to aoip.ai's design is its peer-to-peer (PTP) architecture, a fundamental shift from traditional server-reliant models. This decentralized approach ensures that data transmission occurs directly between users, eliminating the need for centralized intermediate servers. The result is a significant enhancement in data privacy and security, alongside a notable increase in communication efficiency. By adopting PTP decentralization, aoip.ai not only champions user privacy but also lays the groundwork for a more resilient and reliable communication ecosystem.

*Open-Source Transparency:* In a landscape often obscured by proprietary barriers, aoip.ai stands as a beacon of open-source transparency. By granting public access to source code, aoip.ai fosters an environment of collaborative innovation, allowing developers and researchers worldwide to contribute to and enhance its capabilities. Transparency not only accelerates evolution of aoip.ai but also cultivates a community of trust and engagement, ensuring the platform continuously evolves to meet the dynamic needs of its users.

*Speech AI Democratization:* At the forefront of aoip.ai's technological prowess is its integration of state-of-the-art (SOTA) speech AI models. Emphasizing democratization, aoip.ai provides users and developers the liberty to choose and customize AI models that best suit their needs. Whether it's advanced neural network denoisers, transcription, or wake-word detection, aoip.ai empowers users with a toolkit for unparalleled customization and innovation. This democratization of speech AI not only enhances user experience but also paves the way for groundbreaking advancements in VoIP and IoT applications.

In conclusion, aoip.ai represents a paradigm shift in how we perceive and interact with VoIP and IoT technologies. By championing Peer-To-Peer Decentralization, Open-Source Transparency, and Speech AI Democratization, aoip.ai is not just a technological solution; it's a movement towards a more connected, transparent, and empowered digital world.

# 1. Market Overview and Existing Challenges

The landscape of Voice over Internet Protocol (VoIP) and Internet of Things (IoT) applications is rapidly evolving, driven by increasing demand for efficient, reliable, and intelligent communication systems. In this burgeoning market, our startup introduces aoip.ai, an open-source Audio over Internet Protocol (AoIP) Software Development Kit (SDK) integrated with advanced speech AI technologies. This innovative solution is poised to address several critical challenges currently faced in the VoIP and IoT domains.

### *1.1. Data Privacy and Security in Centralized Systems*

A predominant issue in existing VoIP and IoT frameworks is their reliance on centralized servers for data processing and storage. This architecture raises significant concerns regarding data privacy and vulnerability to centralized points of failure. For instance, centralized servers can be susceptible to DDoS attacks, compromising the integrity and availability of communication services. Furthermore, the handling and storage of user data on these servers pose privacy risks, making them a less desirable choice for applications where data confidentiality is paramount.

### *1.2. Inconsistency in Quality of Service*

The quality of service (QoS) in VoIP and IoT applications varies considerably across different platforms, primarily due to differing network conditions and bandwidth limitations. Users often experience issues like jitter, latency, and packet loss, leading to poor audio quality and unreliable connectivity. This inconsistency poses a significant challenge, especially in applications where clear and uninterrupted communication is essential.

### *1.3. Limited Flexibility and Customization in AI Integration*

Most existing VoIP and IoT solutions offer limited flexibility in terms of integrating advanced AI features, such as speech recognition, noise cancellation, and wake-word detection. The proprietary nature of these platforms often restricts the ability to customize or enhance AI functionalities to suit specific application requirements. This limitation hinders innovation and adaptability, particularly for developers looking to build bespoke solutions with specialized AI capabilities.

### *1.4. Transparency and Collaborative Development*

The closed-source nature of many existing solutions stifles collaborative development and innovation. Without access to the underlying source code, external developers and researchers are unable to contribute improvements or adapt the technology for diverse use cases. This lack of transparency and collaboration ultimately slows down the pace of technological advancement in the VoIP and IoT sectors.

In summary, the current VoIP and IoT landscape is characterized by challenges in data privacy, quality consistency, AI integration flexibility, and collaborative innovation. Our solution, aoip.ai, addresses these challenges head-on, leveraging a decentralized, open-source approach combined with state-of-the-art speech AI technologies. This approach not only enhances the quality and reliability of communication services but also fosters a collaborative ecosystem for continuous improvement and innovation.

## 2. Introduction to aoip.ai

In response to the challenges identified in the VoIP and IoT domains, our startup has developed aoip.ai, an innovative solution designed to revolutionize the way audio communication is handled in these sectors. aoip.ai is an open-source Audio over Internet Protocol (AoIP) Software Development Kit (SDK) integrated with advanced speech AI technologies. This section outlines the key aspects of aoip.ai, highlighting its unique features and how it addresses the existing market challenges.

### *2.1. Open-Source Architecture*

At the heart of aoip.ai lies its open-source architecture, a pivotal choice that enables transparency and collaborative development. By providing access to its source code, aoip.ai invites developers and researchers worldwide to contribute to its enhancement and adaptation. This open-source approach not only accelerates technological advancements but also ensures a high degree of adaptability and customization, catering to a wide range of VoIP and IoT applications.

### *2.2. Peer-to-Peer Decentralization*

Diverging from the traditional centralized server model, aoip.ai employs a peer-to-peer (PTP) architecture. This decentralized approach significantly enhances data privacy and security, as there is no central point of data collection or storage. Additionally, PTP decentralization reduces latency and increases the reliability of communications, ensuring a consistent and high-quality user experience across various network conditions.

### *2.3. Advanced Speech AI Integration*

aoip.ai stands out with its integration of cutting-edge speech AI technologies. These technologies include sophisticated neural network-based denoisers, accurate speech recognition systems, and responsive wake-word detection algorithms. By embedding these AI capabilities into the SDK, aoip.ai offers unparalleled improvements in audio quality and intelligent interaction, making it a robust platform for developing advanced VoIP and IoT applications.

*2.4. Customization and Flexibility*

The integration of speech AI in aoip.ai is designed with flexibility in mind. Developers have the freedom to choose, customize, and even enhance AI models based on their specific application requirements. This level of customization is a significant departure from the one-size-fits-all approach commonly seen in existing platforms, allowing for more tailored and effective VoIP and IoT solutions.

*2.5. Addressing Market Needs*

By combining open-source transparency, peer-to-peer decentralization, and advanced speech AI integration, aoip.ai directly addresses the core challenges of data privacy, service quality inconsistency, and limited AI integration flexibility in the current market. This comprehensive approach positions aoip.ai as a forward-thinking solution capable of transforming the VoIP and IoT landscape.

In conclusion, aoip.ai represents a significant leap forward in the development of VoIP and IoT technologies. Its unique combination of open-source architecture, decentralized communication, and advanced AI integration offers a powerful toolkit for developers and a superior experience for end-users, setting a new standard in the industry.

## 3. Peer-to-Peer Decentralization in aoip.ai

A fundamental component of aoip.ai's architecture is its adoption of a peer-to-peer (PTP) decentralized model. This section delves into the technical aspects of this model and its implications for VoIP and IoT applications.

### *3.1. Decentralized Data Transmission*

In traditional VoIP and IoT systems, data typically travels through centralized servers, which can introduce points of failure and security vulnerabilities. In contrast, aoip.ai's PTP architecture ensures that data transmission occurs directly between users. This method not only mitigates the risk of centralized data breaches but also reduces dependency on a single point of failure, thereby enhancing overall system robustness.

### *3.2. Reduced Latency and Improved Scalability*

The decentralized nature of aoip.ai significantly cuts down on the latency usually associated with data routing through central servers. By enabling direct communication paths, the PTP model facilitates a more efficient data transfer, which is critical for real-time audio applications. Additionally, this architecture scales more effectively, as the addition of new nodes (users) does not overload a central server, thereby maintaining system performance even as the user base grows.

### *3.3. Enhanced Data Privacy and Security*

Privacy and security are paramount in communication technologies. Aoip.ai's PTP model inherently bolsters data privacy since the lack of a centralized server reduces the potential for mass data interception or leakage. Furthermore, the architecture allows for the implementation of end-to-end encryption protocols directly between communicating peers, thereby significantly enhancing data security.

### *3.4. Network Resilience and Reliability*

PTP networks are inherently more resilient than centralized systems. In aoip.ai, the decentralized network can adaptively reconfigure in the event of node failures, ensuring continuous service availability. This resilience is crucial for IoT applications where consistent connectivity is vital for device functionality and data integrity.

*3.5. Technical Implementation*

From an engineering perspective, implementing a PTP architecture in aoip.ai involves establishing direct communication channels between nodes, utilizing advanced routing algorithms to optimize data paths, and incorporating robust encryption mechanisms for secure data transfer. This approach requires careful consideration of network topologies, bandwidth management, and latency optimization to ensure a high-quality user experience.

In summary, the peer-to-peer decentralization of aoip.ai offers significant advantages in terms of data privacy, security, latency reduction, scalability, and network resilience. This architectural choice not only addresses the inherent limitations of traditional centralized systems but also aligns with the growing demand for more secure and efficient communication technologies in the VoIP and IoT sectors.

# 4. Open-Source Framework

The open-source nature of aoip.ai is a cornerstone of its design, reflecting a commitment to transparency, collaboration, and continuous improvement in the realms of VoIP and IoT. This section outlines the technical aspects and advantages of aoip.ai's open-source framework.

### *4.1. Collaborative Development and Innovation*

Aoip.ai's open-source model facilitates a collaborative environment where developers and researchers can contribute to and refine the platform. This collaborative development is enabled through publicly accessible repositories, comprehensive documentation, and a community-driven approach to problem-solving and feature enhancement. The collective intelligence of a diverse developer community accelerates innovation and leads to more robust and versatile solutions.

### *4.2. Transparency and Trust*

Transparency is a critical factor in technology adoption, especially in systems handling sensitive data. The open-source framework of aoip.ai ensures complete visibility into the codebase, allowing for thorough security audits and verifications. This transparency builds trust among users and stakeholders, as they can independently assess and verify the security and reliability of the system.

### *4.3. Adaptability and Customization*

The ability to access and modify the source code makes aoip.ai highly adaptable to specific needs. Developers can customize the SDK to suit unique application requirements, whether it's tweaking existing functionalities or integrating new features. This level of customization is particularly beneficial for specialized VoIP and IoT applications that require tailored solutions.

### *4.4. Technical Sustainability and Community Support*

Open-source projects like aoip.ai benefit from ongoing community support and contributions, leading to a sustainable and continuously evolving technology. Regular updates, bug fixes, and feature additions from the community ensure that the platform remains current with the latest technological advancements and industry standards.

### *4.5. Implementation and Quality Assurance*

From a technical implementation standpoint, the open-source framework of aoip.ai involves rigorous code reviews, adherence to coding standards, and comprehensive testing protocols. These practices ensure high-quality code and reliable performance. Furthermore, community feedback and contributions are integral to the iterative improvement process, ensuring that the platform is consistently refined and enhanced.

In conclusion, the open-source framework of aoip.ai is not just a feature but a strategic choice that underpins its technical excellence and future-proofing in the dynamic fields of VoIP and IoT. By embracing an open-source model, aoip.ai leverages the collective expertise of the global developer community, ensuring a platform that is transparent, adaptable, and continuously evolving to meet the ever-changing demands of these industries.

# 5. Integration of Advanced Speech AI in aoip.ai

Aoip.ai represents a significant leap in Audio over Internet Protocol (AoIP) technology by seamlessly integrating advanced speech AI capabilities, particularly through its ESPnet integration. This section elaborates on the scientific and technical nuances of this integration, tailored for an audience deeply involved in research and development, such as the researchers at Carnegie Mellon University.

### *5.1. ESPnet Integration and Model Flexibility*

Aoip.ai incorporates ESPnet, a leading toolkit for end-to-end speech processing, offering users the flexibility to specify and employ their own audio models. This integration empowers researchers and engineers to utilize their custom neural network architectures within AoIP contexts. The focus of aoip.ai is on providing a robust SDK that acts as a conduit for these advanced models, rather than on designing the models themselves. This approach ensures that aoip.ai is adaptable to a wide range of research and application-specific needs.

### *5.2. Technical Synergy with Neural Network Architectures*

The SDK is engineered to be compatible with a variety of neural network architectures, including but not limited to Convolutional Neural Networks (CNNs), Recurrent Neural Networks (RNNs), and Transformer models. This compatibility is crucial for embracing the diversity and rapid evolution of speech AI research. By facilitating the integration of cutting-edge and experimental models, aoip.ai positions itself as a versatile tool for advancing AoIP technologies.

### *5.3. Application in Speech Recognition and Processing*

Aoip.ai's integration with ESPnet particularly shines in applications involving speech recognition, noise reduction, and other audio processing tasks. The flexibility to incorporate models optimized for specific tasks allows for highly accurate and efficient speech recognition capabilities, essential in both VoIP and IoT contexts. This aspect of aoip.ai is particularly relevant for research endeavors focused on improving speech processing algorithms and their practical applications.

*5.4. Enhancement of VoIP and IoT Applications*

The inclusion of advanced speech AI models in aoip.ai significantly enhances the quality and functionality of VoIP and IoT applications. For instance, noise reduction models can provide clearer communication in noisy environments, while wake-word detection algorithms can facilitate more responsive IoT devices. The ability to experiment with and deploy various AI models enables a level of customization and performance tuning previously unattainable in standard AoIP solutions.

*5.5. Scientific Collaboration and Advancement*

Aoip.ai serves as a platform for scientific collaboration and advancement in the field of speech AI. By enabling easy integration of diverse neural network models, it fosters an environment where researchers can experiment, share, and refine their work. This collaborative potential is particularly valuable in academic and research settings, where innovation and experimentation are key drivers of technological progress.

In summary, the integration of advanced speech AI through ESPnet in aoip.ai represents a significant stride in the field of AoIP technology. This feature not only enhances the practical applications of VoIP and IoT systems but also opens up new avenues for scientific research and collaboration in speech processing and neural network architectures.

## 6. Summary and Transition to SDK

Having detailed the core features of aoip.ai — its peer-to-peer decentralization, open-source framework, and integration of advanced speech AI — this section provides a concise summary of these attributes and transitions into a discussion of the SDK architecture, which forms the backbone of aoip.ai's functionality.

### 5.1. Summary of aoip.ai's Core Features

Aoip.ai emerges as a groundbreaking solution in the VoIP and IoT landscape, distinguished by its innovative architecture and advanced functionalities. Its peer-to-peer decentralization addresses critical issues of data privacy and network resilience, setting a new standard in secure and efficient communication. The open-source nature of the platform fosters a collaborative environment for continuous improvement and adaptability, essential in the fast-evolving tech world. Furthermore, the integration of advanced speech AI, particularly through ESPnet, propels aoip.ai to the forefront of technological innovation, offering enhanced audio processing capabilities and a flexible framework for incorporating cutting-edge AI models.

### 5.2. Transition to SDK Architecture

The technical prowess of aoip.ai is encapsulated in its Software Development Kit (SDK), designed to provide a comprehensive and versatile toolkit for VoIP and IoT application development. As we transition to examining the SDK architecture, it is essential to understand how it underpins the functionalities and features highlighted earlier. The SDK is not just a collection of tools but a well-orchestrated framework that enables seamless integration of advanced technologies and customization to meet diverse application needs.

### 5.3. Emphasizing SDK's Role in Implementation

The SDK architecture is the cornerstone of aoip.ai's ability to offer a decentralized, secure, and customizable platform. It is engineered to facilitate easy integration of speech AI models, ensure compatibility with various network architectures, and provide robust tools for developing and deploying VoIP and IoT applications. This section will delve into the technical specifics of the SDK, including its modular design, interoperability with different neural network architectures, and the mechanisms that allow for efficient and secure data transmission in a PTP environment.

## 5.4. Setting the Stage for In-Depth Discussion

In the upcoming section, we will explore the SDK's architecture in detail, focusing on its components, functionalities, and the technical innovations that enable aoip.ai to redefine the standards in VoIP and IoT technologies. This discussion will provide a comprehensive understanding of how aoip.ai's features are practically implemented and the ways in which the SDK empowers developers to build advanced, customized solutions in the field of audio communication.

In conclusion, the combination of aoip.ai's decentralized structure, open-source ethos, and speech AI integration is effectively actualized through its sophisticated SDK. The following section will provide a deeper insight into the SDK architecture, demonstrating how it serves as the technical foundation for realizing the full potential of aoip.ai in various applications.

# 7. Software Developer Kit (SDK) Architecture & Design

The architecture and design of the aoip.ai Software Development Kit (SDK) are crafted to support a range of functionalities vital for advanced VoIP and IoT applications. This section delves into four specific use cases, each illustrating the SDK's capabilities and design considerations. These use cases are especially relevant for understanding aoip.ai's practical implementation and efficiency, and will be presented with technical precision to cater to an audience of software engineers and technical experts.

## 7.1. Static Streaming for Benchmarking

Static streaming for benchmarking is essential for evaluating the performance and robustness of audio AI models under controlled conditions. This approach allows for the fine-tuning of models, ensuring they meet the required standards for accuracy and efficiency, which is vital for maintaining high-quality service in real-world applications.

**Technical Focus**: Static streaming use case focuses on benchmarking the performance of various audio AI models in controlled environments. This is crucial for assessing model efficiency, accuracy, and resource utilization under different network conditions.

**Design Details**: The SDK includes tools for simulating various network environments, logging performance metrics, and analyzing audio quality. It allows for the injection of predefined audio streams into the system to evaluate the performance of AI models in terms of latency, accuracy, and computational load.

Aoip.ai's approach to static streaming for benchmarking stands out due to its comprehensive simulation environment that allows for detailed analysis and optimization of AI models. This method enables a level of precision in performance evaluation that is uncommon in standard VoIP solutions, ensuring that models are rigorously tested and optimized for a variety of network conditions.

A simulation, here, is a process that transmits a given audio signal over voip to generate a relayed audio signal, containing voip + transmitted noise.

Each simulation has a set of configurable parameters as defined below:
```
    [OPTIONAL ] Tracking id                 : a unique id for every simulation
    [REQUIRED] Sender region                : aws region of the sender instance eg: us-east-1
    [REQUIRED] Receiver region              : aws region of the receiver instance
```

```
    [REQUIRED] Sender instance type         : ec2 instance type of the sender instance eg: t2.micro
    [REQUIRED] Receiver instance type       : ec2 instance type of the receiver instance
    [REQUIRED] Sender instance ami          : ami-id of the sender instance eg: ami-0f2b6f057e0b94d5f
    [REQUIRED] Receiver instance ami        : ami-id of the receiver instance
    [REQUIRED] Sender upload bandwidth      : upload bandwidth of the sender instance in kbps eg: 100
    [REQUIRED] Sender download bandwidth    : download bandwidth of the sender instance in kbps
    [REQUIRED] Receiver upload bandwidth    : upload bandwidth of the receiver instance in kbps
    [REQUIRED] Receiver download bandwidth  : download bandwidth of the receiver instance in kbps
    [REQUIRED] src audio config             : audio configuration of the file to be played eg:
test_clean
    [REQUIRED] duration                     : the duration of the file in seconds
    [REQUIRED] s3_bucket_url                : the data source where the recorded file will be stored
```

A simulation can then be run using the utils/simulate_base.py script in the following manner

```
python utils/simulate.py --sreg ap-northeast-2 --rreg ap-northeast-3 --sins
t2.micro --rins t2.micro --sami ami-0c9c942bd7bf113a2 --rami ami-0da13880f921c96a5
--src test_src_noisy --dur 16 --subw 100 --sdbw 100 --rubw 100 --rdbw 100 --s3
s3://raw-src-files/src_noisy_test/
```

The script performs a sequence of steps. It starts by creating and setting up the sender and receiver instances. The processes for both are identical. An instance is specific to a region, as passed in the parameters.It first creates an aws security key and stores it locally. It then creates an aws security group with the required permissions. These permissions are:
1. **TCP Traffic on Port 22:** Allows TCP traffic on port 22 (commonly used for SSH - Secure Shell connections) from any IP address (0.0.0.0/0).
2. **UDP Traffic on Ports 3000 to 9000:** Enables UDP traffic for ports ranging from 3000 to 9000 from any IP address.

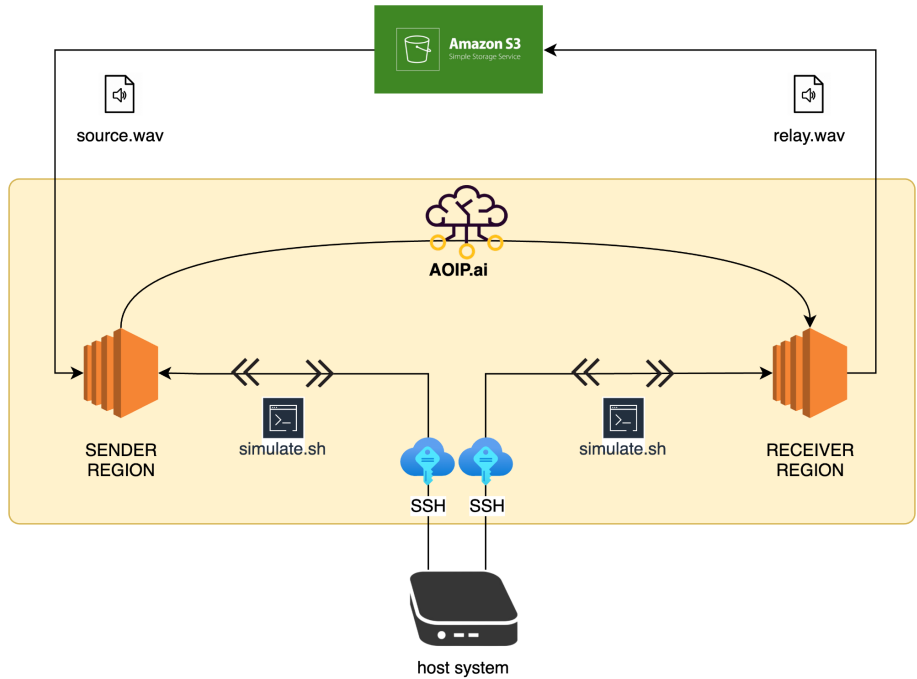

The script then uses the security key, security group, instance type, ami-id and a unique instance name (sender_{tracking_id}) to create an instance. We use a disk size of 8 by default. Once the instance is up and running, we set it up for Aoip.ai to run. We first SSH into the instance using the paramiko library. Then we run a set of commands remotely. The exact commands involved in the setup can be found in config/setup.sh. We essentially install a few libraries: libasound2-dev, sox, ffmpeg, awscli, wondershaper, pulseaudio and then we build our software. The next steps are to pull the source audio to the instance and be ready to relay.

The exact steps are performed on both the sender and receiver instances. The instantiation + setup on average takes 15-17 minutes to complete for a t2.micro. However, since the setup is repetitive and same for each simulation we perform, we build our own custom amis, pre-installed and pre-configured with the setup. These ami-ids can be found in the utils/constants.py file. Thus, we now make use of the utils/simulate.py file. It is a more concise and faster way to run simulations. We have a fully configured instance up and running in 3 minutes or less.

```
python  utils/simulate_ami.py  --sreg  us-east-1  --rreg  us-east-1  --sins  t2.micro
--rins  t2.micro  --sami  ami-0f2b6f057e0b94d5f  --rami  ami-0f2b6f057e0b94d5f  --src
test_src_noisy  --dur  1655  --subw  100  --sdbw  100  --rubw  100  --rdbw  100  --s3
s3://raw-src-files/src_noisy_test_4/
```

The original file, however, still exists. It is more flexible and can be used for generating custom audio simulations. Once both the sender and receiver instances are set up, we put the bandwidth restrictions in place using wondershaper. We now initiate a call.

The receiver instance runs the command

```
cd /home/ubuntu/pjproject-2.13/pjsip-apps/bin && ./pjsua-x86_64-unknown-linux-gnu
--use-cli --auto-answer 200 --auto-rec --rec-file {record_file} --local-port=5061
--ip-addr {receiver_instance_public_ip} --stun-srv=stun.l.google.com:19302
```

We run the pjsua executable with the following params on the receiver side
```
    1. --use-cli          : enables command line control during run
    2. --auto-answer 200  : automatically answer any call that comes through
    3. --auto-rec         : automatically record any call that comes through
    4. --rec-file         : path where the recorded file needs to be stored
    5. --local-port       : the port on which the sip server is running
    6. --ip-addr          : the public ip address of the current machine
    7. --stun-srv         : provide a consistent dns for the ip over long hops
```

The sender instance runs the command

```
cd /home/ubuntu/pjproject-2.13/pjsip-apps/bin && ./pjsua-x86_64-unknown-linux-gnu
--use-cli --auto-play --play-file {src_file} --local-port=5061 --no-vad --ip-addr
{caller_instance_public_ip} --stun-srv=stun.l.google.com:19302
sip:{receiver_instance_public_ip}:5061
```

We run the pjsua executable with the following params on the sender side. A few parameters are common, the others are defined below
```
    1. --auto-play   : automatically play audio on receiving a connection
    2. --play-file   : path of the file to be played
    3. --no-vad      : disable voice activity detection.
                       Ensures continued unaltered audio transmission
    4. sip           : the receiver instance's public ip address.
```

We run these commands async from the script, wrapped in a tmux session to ensure non-blocking and disconnect our ssh session.

```
tmux new-session -d -s {tmux_name} && tmux send-keys -t {tmux_name}
'{pjsua_command}' C-m
```

This starts a new tmux session and send-keys essentially sends the pjsua command to be run. Prolonged ssh connections usually result in disconnects and failures. Once the call begins, we wait for the "duration" amount of time as specified in the input params.

Once the "duration" is over, we ssh into the instances and kill the respective sip sessions. At this point the role of the sender instance is over. We now remove the bandwidth restrictions. As part of the post processing, we process the recorded raw audio on the receiver instance using ffmpeg to a usable format.

```
ffmpeg -i {record_file} -acodec pcm_s16le -ac 1 -ar 16000 {record_ffmpeg}
```

Finally we upload this recorded audio to our s3 bucket url passed in the params. We then terminate our instances and delete the associated security groups and keys.

The simulation is now complete. We constantly log any and all steps involved in the process for efficient debugging in the logging directory. That has been one of the most useful resources that allowed us to reach a scaled completion accuracy of 99.44% (725 success over 729 parallel simulations).

**Setting Up And Scaling aoip.ai**

We heavily use the AWS ecosystem to run our simulations and for storing data. To run the toolkit one needs programmatic access to AWS. Thus, we generate credentials and configure our aws environment with that.

```
AWS_ACCESS_KEY_ID
AWS_SECRET_ACCESS_KEY
```

Apart from that to scale simulations, we need to reset a few default ssh params on linux machines. By default we are limited to 10 ssh connections. That number should be changed to 2x the number of parallel simulations. Say for 500 simulations The exact params to change are:

```
file /etc/ssh/sshd_config

MaxSessions 1000
MaxAuth
MaxStartups 10000:1000:1
```

```
file /etc/ssh/ssh_config

ConnectTimeout 60
```

Then run

```
sudo systemctl restart sshd
sudo systemctl restart ssh
```

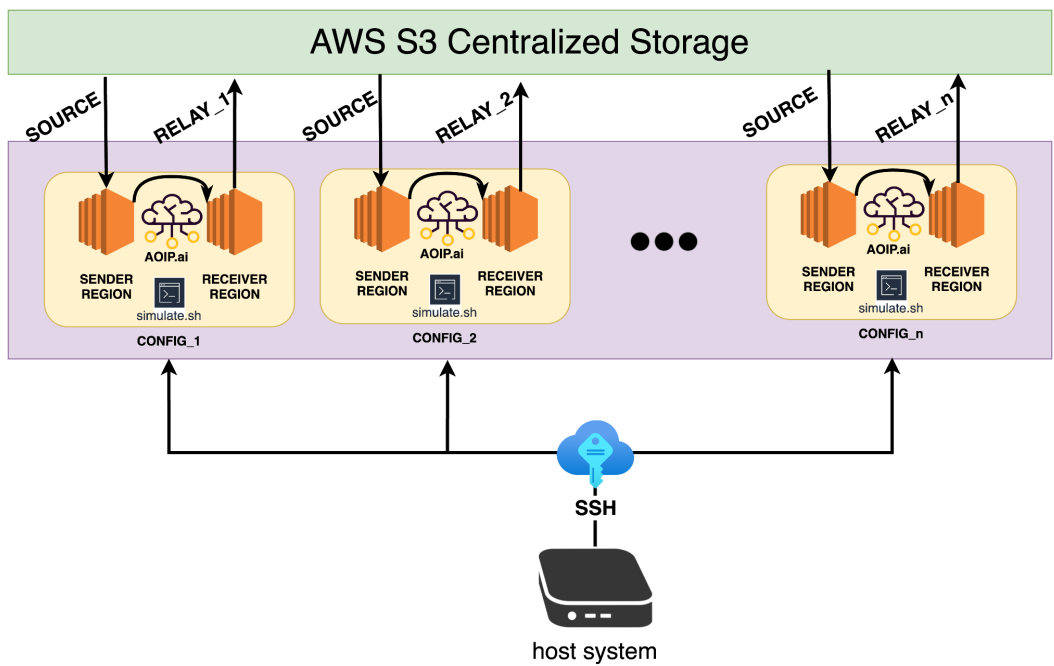

On aws by default you are limited to 100 vCPUs for the on demand instances. Get that limit increased to say 300 or 500. Further, not all regions are enabled by default, please get them enabled too.

## 6.2. Dynamic Streaming for Training

Dynamic streaming for training is key to keeping the AI models adaptive and relevant. By continuously training models with live data, this solution ensures that the AI components of aoip.ai evolve with user behavior and environmental changes, leading to more accurate and responsive AI performance in practical settings.

**Technical Focus**: Dynamic streaming is centered on training and refining AI models. This use case involves streaming live audio data to continuously train and improve AI models, adapting to real-world conditions and user interactions.

**Design Details**: The SDK supports the integration of live data streams with AI training modules. It includes features for data annotation, model retraining, and feedback loops. The design ensures that dynamic data streams are efficiently processed and utilized for incremental learning, maintaining system performance and model accuracy.

The dynamic streaming feature of aoip.ai is unique in its ability to continuously adapt and improve AI models using real-time data. This ongoing training process ensures that the AI components remain cutting-edge and highly responsive to changing user needs and environmental factors. This approach represents a significant advancement in creating adaptive and intelligent communication systems.

The aoip.ai toolkit allows researchers and developers alike to generate an infinite stream of online voip data that can be used for training over online generated data. The toolkit contains a script **utils/dynamic_mixing.py** that reads data from a source directory, relays it over voip and then stores it back to the relay directory. Here is how to run this:

```
python utils/dynamic_mixing.py --src_dir src_noisy --relay_dir  src_noisy --clean_dir src_clean --num 100
```

Below we define the params:
```
   1. --src_dir    : the directory that contains the source data
   2. --relay_dir  : the directory that contains the relay data
   3. --clean_dir  : the directory that contains the parallel clean data
                     (before relay)
   4. --num        : number of simulations that run in parallel
```

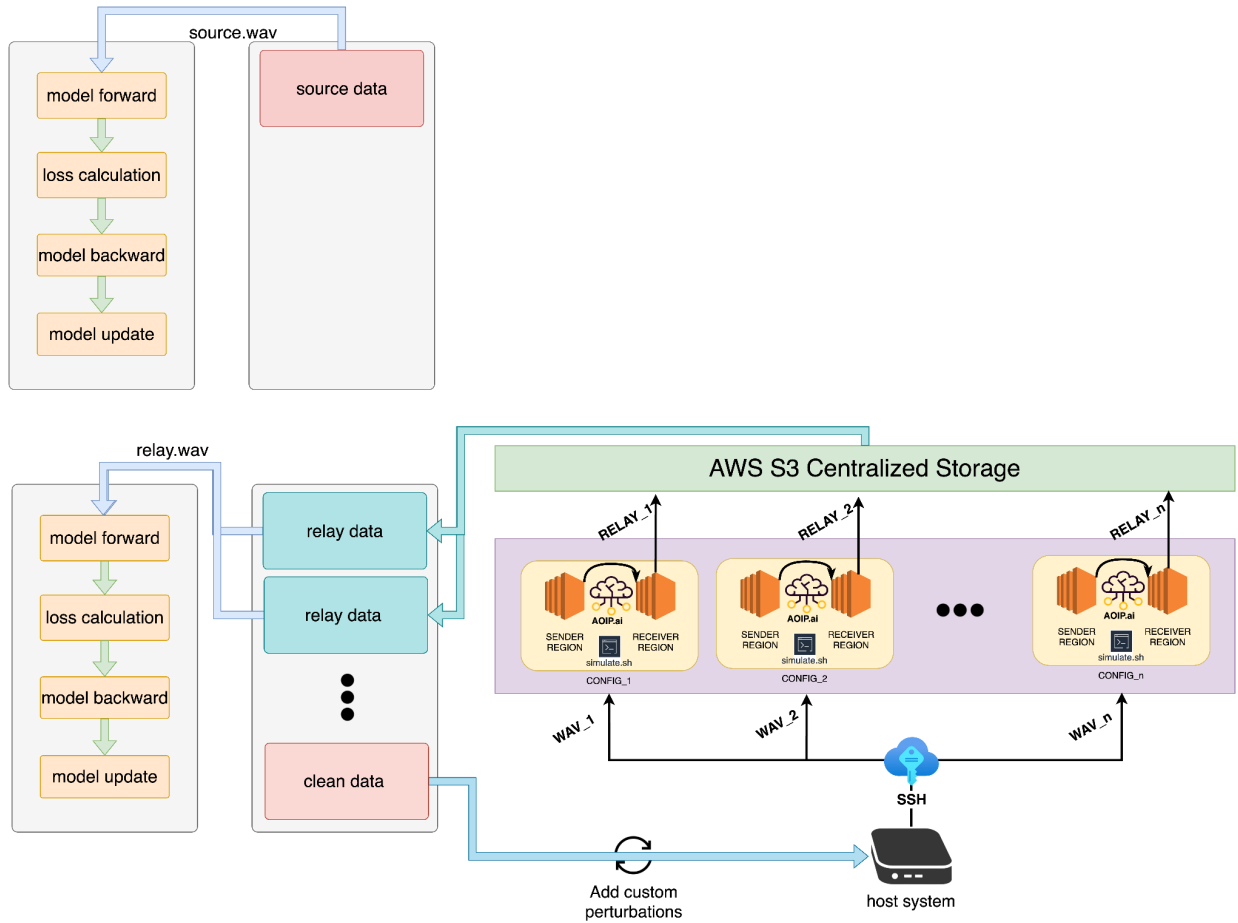

Let's look into the process. We first initialize the number of simulations as passed by the num parameter. A simulation here is defined differently in the sense that it is not terminated after relaying just one audio. It will continue to run. Once we have them up and running, we begin the relaying process. The dataloader reads the file names into memory and every file is considered one job. We use the ThreadPoolExecutor library from python. Each job is assigned to a simulation. It then relays it over to the receiver, the relayed audio is sent back to the source machine and the simulation is now free to do the next job. The thread pool executor executes 'num' number of jobs in parallel and continues to assign subsequent jobs to any simulation that gets free. Once all jobs are finished the instances are terminated and all resources are deleted. We also share a notebook to show how dynamic mixing works in the background while training demucs.

Thus, this dynamic mixer can run in the background while model training is in progress. We can continue to add any perturbations or special effects to the audio before relay to have an infinite stream of source data. We continue to log any and all steps involved in the process for efficient debugging in the logging directory.

## 6.3. Comprehensive Evaluation Methodology

A comprehensive evaluation methodology is indispensable for ensuring the overall quality and effectiveness of the aoip.ai system. This holistic approach to testing and evaluation guarantees that every aspect of the system, from network performance to user experience, adheres to the highest standards, ensuring reliability and user satisfaction.

**Technical Focus**: This use case encompasses a thorough evaluation methodology for the entire AoIP system, including AI models, network performance, and end-user experience.

**Design Details**: The SDK provides a suite of evaluation tools for comprehensive testing, including automated test scripts, performance monitoring tools, and user experience surveys. This methodology involves stress testing under various network conditions, user feedback analysis, and AI model performance evaluation. The aim is to ensure that the SDK meets high standards of reliability, scalability, and usability.

Aoip.ai's comprehensive evaluation methodology is distinctive in its holistic and rigorous approach to system testing. By encompassing all aspects of the system, including network performance, AI model efficiency, and user experience, this methodology ensures a thorough and balanced evaluation. This extensive testing framework is key to delivering a consistently high-quality and reliable solution.

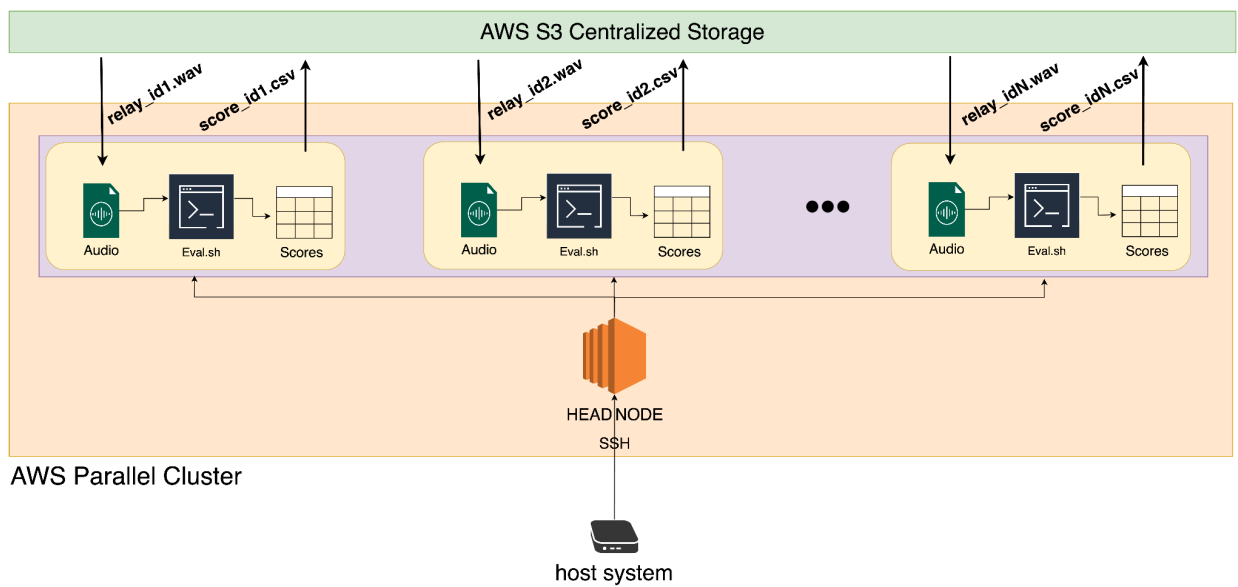